\documentclass[fleqn,usenatbib]{mnras}

\usepackage{newtxtext,newtxmath}

\usepackage[T1]{fontenc}
\usepackage{ae,aecompl}

\usepackage{graphicx}	
\usepackage{amsmath}	
\usepackage{amssymb}	

\usepackage{lineno}

\title[Polarimetric constraint on jet emission of Cyg X-1]{${\it PoGO+}$ polarimetric constraint on the synchrotron jet emission of Cygnus X-1}

\author[M. Chauvin et al.]{
Maxime Chauvin,$^{1,2}$
Hans-Gustav Flor\'{e}n,$^{3}$
Miranda Jackson,$^{1,4}$
Tuneyoshi Kamae,$^{5,6}$
\newauthor
Jun Kataoka,$^{7}$
M\'{o}zsi Kiss,$^{1,2}$
Victor Mikhalev,$^{1,2}$
Tsunefumi Mizuno,$^{8}$
\newauthor
Hiromitsu Takahashi,$^{8}$\thanks{E-mail: 
hirotaka@astro.hiroshima-u.ac.jp (HT)}
Nagomi Uchida$^{8}$
and Mark Pearce$^{1,2}$
\\
\\
$^{1}$Department of Physics, KTH Royal Institute of Technology, 106 91 Stockholm, Sweden\\
$^{2}$The Oskar Klein Centre for Cosmoparticle Physics, AlbaNova University Centre, 106 91 Stockholm, Sweden\\
$^{3}$Department of Astronomy, Stockholm University, 106 91 Stockholm, Sweden\\
$^{4}$School of Physics and Astronomy, Cardiff University, Cardiff CF24 3AA, UK\\
$^{5}$Department of Physics, University of Tokyo, Tokyo 113-0033 Tokyo, Japan\\
$^{6}$SLAC/KIPAC, Stanford University, 2575 Sand Hill Road, Menlo Park, CA 94025, USA\\
$^{7}$Research Institute for Science and Engineering, Waseda University, Tokyo 169-8555, Japan\\
$^{8}$Department of Physical Science, Hiroshima University, Hiroshima 739-8526, Japan\\
}

\date{Accepted XXX. Received YYY; in original form ZZZ}

\pubyear{2018}

\begin{document}
\label{firstpage}
\pagerange{\pageref{firstpage}--\pageref{lastpage}}
\maketitle

\begin{abstract}
We report a polarimetric constraint on the hard X-ray synchrotron jet emission from the \mbox{Cygnus X-1} black-hole binary system.
The observational data were obtained using the ${\it PoGO+}$ hard X-ray polarimeter in July 2016, when \mbox{Cygnus X-1} was in the hard state.
We have previously reported that emission from an extended corona with a low polarization fraction is dominating, and that the polarization angle is perpendicular to the disk surface.
In the soft gamma-ray \textcolor{black}{regime}, a highly-polarized synchrotron jet is reported with ${\it INTEGRAL}$ observations.
To constrain the polarization fraction and flux of such a jet component
in the hard X-ray \textcolor{black}{regime},
we now extend analyses through vector calculations in the Stokes $QU$ plane,
where the dominant corona emission and the jet component are considered simultaneously.
The presence of another emission component with different polarization angle could partly cancel out the net polarization.
The 90\% upper limit of the polarization fraction for the additional synchrotron jet component is estimated as $<$10\%, $<$5\%, and $<$5\% for polarization angle perpendicular to the disk surface, parallel to the surface, and aligned with the emission reported by ${\it INTEGRAL}$ data, respectively.
From the 20--180~keV total flux of \mbox{2.6 $\times$ 10$^{-8}$ erg s$^{-1}$ cm$^{-2}$}, the upper limit of the polarized flux is estimated as \mbox{$<$ 3 $\times$ 10$^{-9}$ erg s$^{-1}$ cm$^{-2}$}.
\end{abstract}



\nolinenumbers

\section{Introduction}
\label{sec:intro}

Black-hole binaries (BHBs) consist of a stellar-mass black hole (BH) and a companion star.
The BH accretes matter from the star, thus forming an accretion disk, corona and jet structures.
Although the existence of a jet in a BHB is confirmed observationally \citep[for a review]{jet_fender}, the underlying physics (e.g., energetics and formation mechanism) are not yet understood in detail.
There are two jet types.
One is a 'transient' jet, associated with state transitions, and its image is sometimes directly resolvable by radio interferometry.
It exhibits superluminal motion and is accelerated close to the speed of light \citep[e.g., ][]{jet_grs1915, jet_groj1655}.
The other is a 'compact' jet present in the quiescent and hard states.
Although not resolved by radio images, the radio emission, which is considered to arise from a self-absorbed synchrotron jet, has a hard spectral index and an infrared flux exceeding \textcolor{black}{that} is extrapolated from black-body emission of the accretion disk \citep[e.g., ][]{jet_gx339}.

Cygnus X-1 (Cyg X-1) is one of the persistently bright BHBs in our Galaxy \citep{cygx1_bh}.
It is predominantly in the hard state, where the spectrum has a hard spectral index of $\sim$1.7 and a peak around \mbox{100 keV},
\textcolor{black}{i.e.,} suitable to study the jet physics.
Cyg X-1 is a high-mass X-ray binary
with a BH of mass $(15\pm1) M_{\odot}$ and
a supergiant companion star \citep{cygx1_mass1, cygx1_mass2}.
The jet structure of \mbox{Cyg X-1} in the hard state was resolved  with radio images \citep{cygx1_radio1,cygx1_radio2}.
Mid-infrared spectral and near-infrared and optical polarimetric observations are described by synchrotron jet emission \citep{cygx1_mir, cygx1_ir_opt}.
Using soft gamma-ray spectral and polarimetric information from the ${\it INTEGRAL}$ satellite, the power-law emission above $\sim$300 keV is reported to be highly polarized, $>$75\%, which is also ascribed to the synchrotron jet \citep{integral_ibis, integral_spi}.
The GeV gamma-ray spectrum has been observed by the ${\it Fermi}$ satellite. To explain the multi-wavelength spectral energy distribution (SED), the gamma-ray emission is proposed to arise from the inverse-Compton mechanism by high-energy electrons in the jet \citep{cygx1_fermi}.
If the jet produces strong synchrotron emission above several hundred keV, the strength of the jet magnetic fields is predicted to exceed the equipartition level \citep{cygx1_sed}.
Although the jet is resolved in the radio domain, its flux is relatively low in the higher energy band compared to the companion star, disk and corona in optical, soft X-rays and hard X-rays \citep{cygx1_ir_opt, cygx1_sed}.
The jet SED of \mbox{Cyg X-1} is not yet well understood.

Recently, we used the balloon-borne ${\it PoGO+}$ telescope \citep{PoGO+_Gal} to observe the hard X-ray (19--181 keV) linear polarization of \mbox{Cyg X-1} in the hard state. This energy range is suitable to study the corona emission reflected by the disk.
We have previously shown that the corona geometry is extended rather than compact \citep{PoGO+_CygX1}.
Such discrimination has not been possible 
from previous observations in soft X-rays or soft gamma-rays \citep{cygx1_xray,integral_ibis,integral_spi}.

In this paper, we extend polarization analyses through vector calculations in the Cartesian Stokes $QU$ plane to constrain the flux of an additional highly-polarized synchrotron jet component, as suggested by previous ${\it INTEGRAL}$ observations.
We introduce the ${\it PoGO+}$ observations and polarimetric results \textcolor{black}{at} other wavelengths and numerical simulations in $\S$~\ref{sec:obs}.
We start from analyses \textcolor{black}{where} the major source of hard X-rays from Cygnus X-1 is the extended corona in $\S$~\ref{sec:single}.
In $\S$~\ref{sec:extend}, we add \textcolor{black}{a} possible contribution of the synchrotron emitting jet to the polarization, and set a limit to the contribution.
A low polarization fraction, $PF$, observed from a source can result either from a low intrinsic source polarization, 
or from \textcolor{black}{cancellation} by an additional flux component at polarization angle, $PA$, different from the main flux.
In $\S$~\ref{sec:lamp}, we confirm that the compact corona model predicts a high polarization fraction which is inconsistent with the ${\it PoGO+}$ results, even when considering such a separate emission component.
We \textcolor{black}{present} our conclusions in $\S$~\ref{sec:conclusion}.

\if
A low polarization fraction, $PF$, observed from a source can result either from a low intrinsic source polarization, 
or from destructive interference by an additional flux component at  polarization angle, $PA$, different from the main flux.
In this paper, we assume a dominant corona emission and use vector analysis in the Cartesian Stokes $QU$ plane to constrain the flux of an additional highly-polarized synchrotron jet component, as suggested by previous ${\it INTEGRAL}$ observations.
\fi

\section{Observations and Data Analyses}
\label{sec:obs}

${\it PoGO+}$ is a hard X-ray polarimeter which performed \mbox{Cyg X-1} observations in the 19--181 keV range (median energy of 57 keV) during July 12--18 in 2016 \citep{PoGO+_CygX1}.
The source was in the typical hard state, based on light curves by the ${\it MAXI}$ \citep{maxi} and ${\it Swift}$/BAT \citep{bat} instruments.
Using a previous ${\it Suzaku}$ observation at similar ${\it MAXI}$ and ${\it Swift}$ fluxes \citep{suzaku}, the 20--180 keV flux is estimated as \mbox{$2.6 \times 10^{-8}$ erg s$^{-1}$ cm$^{-2}$} \citep{PoGO+_CygX1}.
At the distance of 1.86 kpc, the luminosity is \mbox{$1.1 \times 10^{37}$ erg s$^{-1}$}, which corresponds to 0.6\% of the 15$M_{\odot}$ Eddington luminosity.

Results from the ${\it PoGO+}$ observation of \mbox{Cyg X-1} give a 'Maximum A Posteriori' (MAP) estimate of \mbox{$PF$ = 4.8\%} and \mbox{$PA$ = 154$^{\circ}$}, where $PA$ is measured from North to East (i.e., counter-clockwise on the sky).
\textcolor{black}{This $PA$ value is consistent with a direction perpendicular to the disk surface.}
Marginalizing the posterior yields \mbox{$PF$ = $(0.0^{+5.6}_{-0.0})$\%} and \mbox{$PA$ = (154 $\pm$ 31)$^{\circ}$}, where marginalized values are obtained by projecting the density map onto the $PF$ and $PA$ axis, respectively.
The point-estimate and the uncertainty correspond to the peak and the region of highest posterior density containing 68.3\% probability content, respectively. Details of the ${\it PoGO+}$ polarization analysis are described in \citet{PoGO+_CygX1}. 

To measure linear polarization, ${\it PoGO+}$ utilizes the anisotropy of azimuthal Compton scattering events, as described by the Klein-Nishina relationship. X-rays are more likely to scatter in the direction perpendicular to the polarization, resulting in a sinusoidal modulation curve with a 180$^{\circ}$ period in the distribution of possible scattering angles \mbox{(0--360$^{\circ}$)}.
Results can be transformed to the Stokes $QU$ plane using the following relations between $PF$, $PA$ and $Q$, $U$:
\begin{equation}
    PF = \sqrt{\left( \frac{Q}{I} \right)^2 + \left( \frac{U}{I} \right)^2}
	\label{eq:PF}
\end{equation}
and
\begin{equation}
	PA = \psi / 2 ,
	\label{eq:PA}
\end{equation}
\textcolor{black}{where $Q$ and $U$ are the fractions of the total intensity $I$ parallel and perpendicular to a specific reference direction, respectively, and}
$\psi$ value is the angle from the positive $Q$ axis in the $QU$ plane
measured in the counter-clockwise direction. Only a linear polarization fraction is considered (no circular polarization component, i.e., Stokes $V=0$). 
In this representation, the distance from the origin is equivalent to the polarization fraction, while the angle~$\psi$ corresponds to twice the polarization angle.
A consequence is that maxima and minima are separated by \mbox{$PA$ = 90$^\circ$} in the modulation curve, corresponding to $\psi=180^\circ$ in the Stokes $QU$ plane.
\textcolor{black}{In the Stokes $QU$ plane, two incoherent polarized components add as vectors, yielding the total $PF$ and $PA$ values observed.}
In X-ray polarization analyses with a sinusoidal modulation curve, observed 'detector Stokes parameters'
$Q/I$ and $U/I$ are limited to the range 0--0.5, due to the \mbox{180$^{\circ}$} period of the modulation curve. Multiplication by a factor of~2 is required to obtain the true 'source Stokes parameters' $Q/I$ and $U/I$ of Eq.~\ref{eq:PF} from 
the definitions
in \citet{stokes_kislat,stokes_victor}.
We adopt the source Stokes formalism in this paper.

\begin{figure}
	\includegraphics[width=\columnwidth]{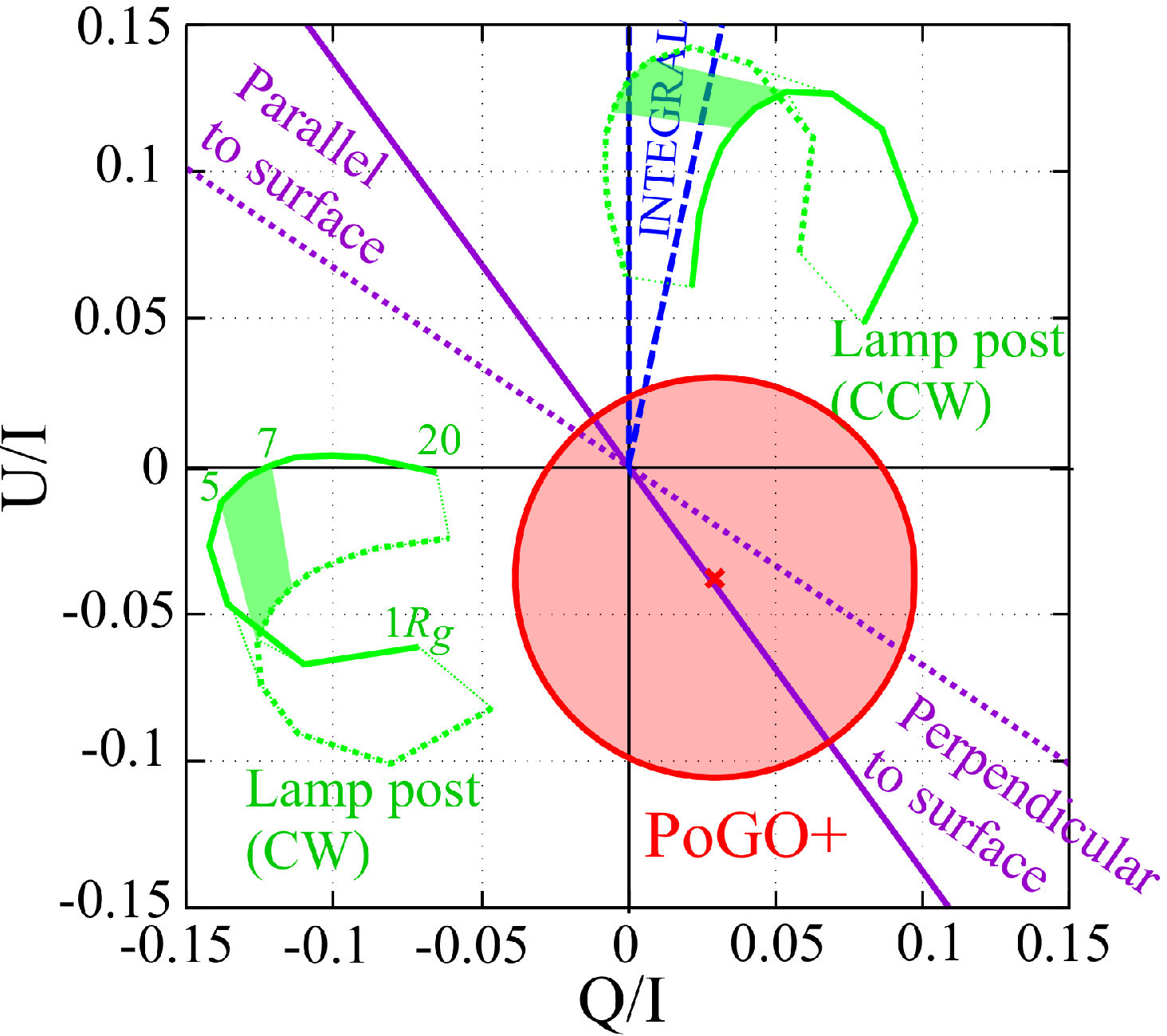}
    \caption{${\it PoGO+}$ hard X-ray polarization results of \mbox{Cyg X-1} in the Stokes $QU$ plane. The 90\% upper limit corresponds to the circle (red), and its center (red cross) is the MAP estimate (\mbox{$PF$ = 4.8\%} and \mbox{$PA$ = 154$^{\circ}$}) \citep{PoGO+_CygX1}.
    Directions parallel and perpendicular to the surface of the accretion disk are in the 2nd and 4th quadrant, respectively (purple).
    These span a range of \mbox{$PA$ $\pm$ $5^{\circ}$} (i.e., \mbox{$\psi\pm10^{\circ}$} in the $QU$ plane) based on the direction changes of the observed radio jet, which we assume to be  perpendicular to the disk surface.
    \textcolor{black}{Solid and dotted lines correspond to $PA$ $-5^{\circ}$ and $+5^{\circ}$, respectively.}
    The $PF/PA$ range predicted for the lamp-post corona model of \mbox{Cyg X-1} is plotted (green) \citep{sim_lamp}.
    The region is defined by
    \textcolor{black}{
    the corona height varying \mbox{1$-$20 $R_g$},
    with the jet direction varying \mbox{$\pm$5$^{\circ}$}
    where solid and dotted lines again correspond to $PA$ $-5^{\circ}$ and $+5^{\circ}$, respectively.
    Corona heights of 1, 5, 7 and 20 $R_g$ are indicated for the $-5^{\circ}$ case.
    Two filled regions are at a corona height 5--7 $R_g$ as estimated from spectral analyses \citep{cygx1_fabian}.}
    The $PA$ direction $(42\pm3)^{\circ}$ of the power-law emission in the several 100~keV range from ${\it INTEGRAL}$ observations is shown in the 1st quadrant (blue) \citep{integral_spi}. As the ${\it INTEGRAL}$ energy range is higher than that of ${\it PoGO+}$ (median 57~keV), they cannot be compared directly, hence the dashed lines.
    }
    \label{fig:QU}
\end{figure}

Fig.~\ref{fig:QU} shows the results from Fig.~2 of \citet{PoGO+_CygX1} represented in the Stokes $QU$ plane. The red cross is the MAP estimate and the red circle corresponds to the 90\% credibility region. It is obtained by taking pairs of $PF$ and $PA$ values along the 90\% credibility contour and mapping them onto the $QU$ plane.
During observations, the ${\it PoGO+}$ polarimeter rotates \mbox{$\pm180^{\circ}$ at $1^{\circ}$ s$^{-1}$} to eliminate instrumental bias.
The rotation makes the detector response flat, with a systematic polarization \mbox{$PF$ = (0.10 $\pm$ 0.12)\%} for non-polarized inputs \citep{PoGO+_Cal}, generating a circular credibility region in the $QU$ plane.
This region is found to have a radius \mbox{$PF$ = 6.9\%}.

Table~\ref{tab:angle} summarizes polarimetric information for \mbox{Cyg X-1} as measured in different wavelengths ($\S$~\ref{sec:intro}) and simulated for the hard X-ray emission.
Measurement bands below the hard X-ray range \mbox{($<200$~keV)} show $PA$ nearly aligned with the radio jet direction, \mbox{(158 $\pm$ 5)$^{\circ}$} \citep{cygx1_radio1, cygx1_radio2}, assumed to be perpendicular to the accretion disk surface.

\begin{table}
	\centering
	\caption{A list of the polarization \textcolor{black}{parameters} ($PF$ and $PA$) in the hard state of Cyg X-1 from previous studies. Observational errors are 1$\sigma$ confidence level, unless stated otherwise.
	$PA$ values of simulations assume that the disk rotation axis is parallel with the radio jet direction $(158\pm5)^{\circ}$
	\textcolor{black}{and include the 
	uncertainty from the radio observations} \citep{cygx1_radio1, cygx1_radio2}.
	$^{a}$ Infrared and optical observations include interstellar polarization effects.
	$^{b}$ Some fraction of the X-ray observations includes data in the soft state but the contribution is not significant.
	$^{c}$ For 3$\sigma$ confidence level.
	$^{d,e}$ See text for details of the simulation setups.
	$^{d}$ For 20--100 keV energy band.
	$^{e}$ For 20--50 keV.
	References (1) \citet{cygx1_radio1}; (2) \citet{cygx1_ir_opt}; (3) \citet{cygx1_opt}; (4) \citet{cygx1_xray}; (5) \citet{PoGO+_CygX1}; (6) \citet{integral_spi}; (7) \citet{sim_extend}; (8) \citet{sim_lamp}.}
	\label{tab:angle}
	\begin{tabular}{lccr}
		\hline
		  & $PF$ (\%) & $PA$ ($^{\circ}$) & Ref \\
		\hline
		\multicolumn{4}{l}{Observations in several bands} \\
		\hline
		5 GHz & $<10$ & --- & 1\\
		1.25--2.16 $\mu$m$^{a}$ & 0.84--1.95 & 136.1--142.8 & 2\\
		0.4--0.9 $\mu$m$^{a}$ & 3.3--5.0 & 136.4--137.6 & 3\\
		2.6 keV$^{b}$ & $2.44\pm1.07$ & $162\pm13$ & 4\\
		5.2 keV$^{b}$ & $5.3\pm2.5$ & $155\pm14$ & 4\\
		19--181 keV & $<5.6$ & $154\pm31$ & 5\\
		130--230 keV & $<20^{c}$ & --- & 6\\
		230--850 keV & $76\pm15$ & $42\pm3$ & 6\\
		\hline
		\multicolumn{4}{l}{Simulations in hard X-rays} \\
		\hline
		Extended Corona$^{d}$ & 2.5 & $158\pm5$ & 7\\
		Lamp-post Corona (CW)$^{e}$ & 9--15 & $103\pm15$ & 8\\
		Lamp-post Corona (CCW)$^{e}$ & 9--15 &
		$33\pm15$ & 8\\
		\hline
	\end{tabular}
\end{table}

\section{Results and Discussions}
\label{sec:result}

\subsection{Single emission component}
\label{sec:single}

In \citet{PoGO+_CygX1}, assuming only one emission component, we obtained the $PF$ upper limit at 90\% confidence level as $<$8.6\% for the corona emission, by marginalizing over the full $PA$ range of \mbox{0--180$^{\circ}$}.
When any $PA$ is allowed, we can determine the 90\% $PF$ upper limit to be as large as 11.6\% (maximum length from the origin to any point on the red circle in Fig.~\ref{fig:QU}).
This occurs when \mbox{$PA$ = 154$^{\circ}$} (i.e., \mbox{$\psi$ = 308$^{\circ}$} in the $QU$ plane) and corresponds to a direction perpendicular to the accretion disk surface.
Similarly, if we consider the emission with \textcolor{black}{a} $PA$ direction parallel to the disk surface or aligned with the highly-polarized power-law emission observed in the several \mbox{100 keV} range, the $PF$ value cannot exceed 2.2\% or 2.9\%, respectively (intersections between the red circle and the purple region in the 2nd quadrant or the blue dashed region in the 1st quadrant of Fig.~\ref{fig:QU}).

Based on previous spectral and timing analyses \citep[e.g., ][]{corona_done}, the hard X-ray emission of BHBs is dominated by a high-temperature corona emission including its reflection off the accretion disk.
There are two main competing models for the corona geometry in the hard state: the extended corona model \citep[e.g., ][]{corona_extend} and the lamp-post corona model \citep[e.g., ][]{corona_lamp}.
The former assumes a larger corona size, with the disk being truncated before reaching the innermost stable circular orbit  \citep[e.g., ][]{cygx1_makishima}.
In the latter model, the corona is assumed to be compact in size and  located on the rotation axis of the black hole, close to the event horizon. Emission near the black hole is influenced by strong relativistic effects \citep[e.g., ][]{cygx1_fabian}.
We concluded that the \textcolor{black}{simple} lamp-post corona model 
was 
not consistent with ${\it PoGO+}$ polarization measurements and that the extended corona model 
was
favored 
instead
\citep{PoGO+_CygX1}.
\textcolor{black}{Other corona models have been proposed (slab, patchy, outflowing, etc.) \citep[e.g.,][]{cygx1_nowak} although the two main models which we considered can be seen as representative of these.}

The extended corona model has a small fraction of the reflection component in the hard X-ray band, and it assumes a lower $PF$ value ($\sim$2.5\%) with $PA$ perpendicular to the disk surface by numerical simulations \citep{sim_extend}.
Therefore, the ${\it PoGO+}$ upper limit of 11.6\% with this $PA$ direction is consistent with the extended corona model.


Conversely, as shown in Fig.~\ref{fig:QU} and Table~\ref{tab:angle}, the lamp-post corona model predicts higher $PF$ (9--15\%) and $PA$ rotation \mbox{(55 $\pm$ 10)$^{\circ}$} relative to the disk rotation axis due to the strongly enhanced reflection emission \citep{sim_lamp}.
\textcolor{black}{We assume a corona height \mbox{1--20 $R_{g}$}, 
the extreme Kerr case and disk-inclination angle 30$^{\circ}$,
following X-ray spectral analyses \citep[e.g.,][]{cygx1_fabian} and the orbital-inclination angle measured in radio \citep{cygx1_inc_radio}.
Here, \mbox{$R_{g}=GM/c^{2}$} is the gravitational radius, with $G$ being the gravitational constant, $M$ the BH mass and $c$ the speed of light.
If the inclination angle is 40$^{\circ}$ as reported from X-ray analyses \citep{cygx1_walton}, the simulated $PA$ rotates more \citep{sim_lamp} and even separates from the observed MAP $PA$ of 154$^{\circ}$.}
From radio observations, the direction of the orbital rotation is estimated to be clockwise (CW) \citep{cygx1_inc_radio}.
If we assume CW rotation (same direction for the accretion disk as for the orbit), the \mbox{55$^{\circ}$} rotation is subtracted from the $PA$, yielding 
\mbox{(103 $\pm$ 15)$^{\circ}$}. For the counter clockwise (CCW) case, the rotation is instead added, resulting in 
\mbox{$PA$ = (33 $\pm$ 15)$^{\circ}$}.
In both the CW and CCW case, the resulting $PA$ differs from the ${\it PoGO+}$ measurement.
Then, the largest possible $PF$ upper limit of 7.6\%,
\textcolor{black}{at $\psi = 236^{\circ} (PA = 118^{\circ})$}
in the 3rd quadrant of Fig.~\ref{fig:QU},
becomes incompatible with the predicted level of 9--15\%
\textcolor{black}{
(i.e., the entire 1--20 $R_g$ corona height range is incompatible with the PoGO+ data).
}

\subsection{Extended corona emission with a synchrotron jet component}
\label{sec:extend}

The limits \textcolor{black}{presented in the previous section} derive from only assuming one emission component.
As described in $\S$~\ref{sec:intro}, polarization observations measure only the total $PF$ and $PA$ (i.e., the summed Stokes vector of the underlying emission components).
In the following, we instead examine a situation where the dominant extended corona emission is complemented by a possible synchrotron jet contribution, as suggested by \citet{integral_ibis} and \citet{integral_spi}.

Although there is no unified picture for the magnetic field in the jet structure, the $PA$ direction is typically assumed to be parallel or perpendicular to the disk rotation axis \citep{intro_jet}.
It is proposed, from infrared and optical observations summarized in Table~\ref{tab:angle}, that the $PA$ of the synchrotron jet emission can be perpendicular to the disk surface \citep{cygx1_ir_opt}.
This would correspond to an upper limit on the total emission $PF_{\rm total}$ of 11.6\%, arising from Fig.~\ref{fig:QU} as described previously.
Since both the extended corona and the additional component have the same $PA$ direction and the extended corona is predicted to have $PF_{\rm corona}$ of a few percent,
the additional synchrotron jet can have \mbox{$PF_{\rm jet}$ $\lesssim$ 10\%},
such that $PF_{\rm corona}$ + $PF_{\rm jet}$ cannot exceed 11.6\%.
In this case, the \mbox{20--180~keV} polarized flux of the jet emission is calculated as \mbox{$<$10\%} of the total flux, 
i.e., \mbox{$<$ 3 $\times$ 10$^{-9}$ erg s$^{-1}$ cm$^{-2}$}.
\textcolor{black}{Here, we define $PF_{\rm corona}$ and $PF_{\rm jet}$ with respect to the total flux.
Simulations assume only corona emission, and we ignore the small change of $PF_{\rm corona}$ due to the contribution of the jet component, as the corona emission is assumed to be dominant throughout this paper.
}

To derive the actual jet flux ($F_{\rm jet}$) from the polarized flux, we need to estimate the $PF$ value of the jet emission at the source.
If we assume that the jet emission is 100\% polarized, $F_{\rm jet}$ is the same as the polarized flux of $<3 \times$ 10$^{-9}$ erg s$^{-1}$ cm$^{-2}$.
However, if the jet emission has only 10\% polarization, which is a typical magnitude for blazar synchrotron jets observed in the optical range \citep{blazar_pol},
$F_{\rm jet}$ becomes a factor of (1/0.1) higher than the polarized flux, i.e., $<3 \times$ 10$^{-8}$ erg s$^{-1}$ cm$^{-2}$.
In this situation, $F_{\rm jet}$ is the dominant component of $F_{\rm total}$, which is inconsistent with the physical picture that the extended corona emission ($F_{\rm corona}$) dominates.


If the additional jet emission has $PA$ parallel with the disk surface, the 90\% upper limit of $PF_{\rm total}$ is lower, with the lowest upper limit being 2.2\%, as mentioned above.
However, in this case, the $PA$ directions are opposite for the extended corona (perpendicular to the disk surface) and the additional component (parallel with disk surface), and 
$PF_{\rm jet}$ can be as high as $\sim$5\%,
where $PF_{\rm total}$ ($<$2.2\%) is obtained as \mbox{$PF_{\rm jet}$ $-$ $PF_{\rm corona}$} 
(\mbox{$PF_{\rm corona}$ = 2.5\%} from Table~\ref{tab:angle}).

\begin{figure}
	\includegraphics[width=\columnwidth]{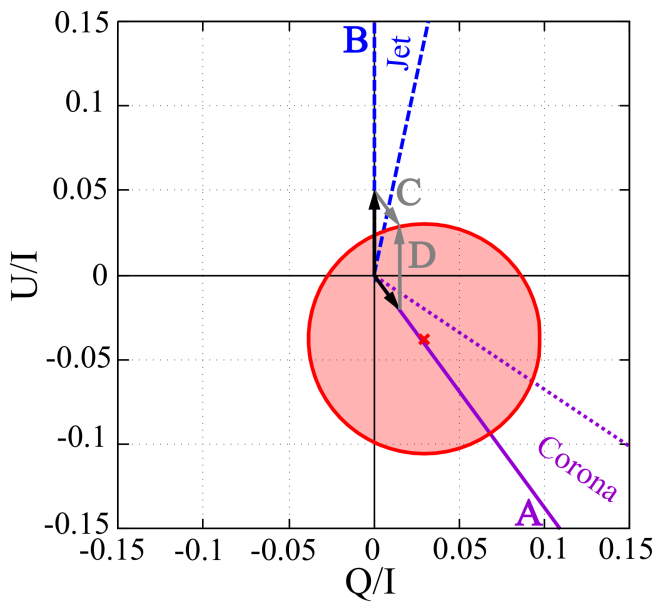}
    \caption{
    Stokes $QU$ vector calculations for the extended corona component and a synchrotron jet component following from the ${\it INTEGRAL}$ SPI data. Lines A and B, for the extended corona and jet emission, respectively, result in the largest angular separation (maximum cancellation).
    For a typical value \mbox{$PF_{\rm {corona}}$ = 2.5\%} (vector~C parallel with line~A), $PF_{\rm {jet}}$ is constrained to below 5\% (vector~D parallel with line B).
    }
    \label{fig:QU_extend}
\end{figure}

We now turn to the case where the additional power-law emission has \mbox{$PA$ = (42 $\pm$ 3)$^{\circ}$}, as suggested by ${\it INTEGRAL}$ SPI measurements for the 230--850 keV region \citep{integral_spi}.
Polarization results from ${\it INTEGRAL}$ IBIS are consistent but have larger errors \citep{integral_ibis, integral_ibis2} and are not considered here.
We consider possible combinations of two Stokes vectors yielding a vector sum within the 90\% upper-limit circle of the ${\it PoGO+}$ measurement.
Fig.~\ref{fig:QU_extend} illustrates this vector addition, where regions follow from Fig.~\ref{fig:QU} but text labels have been removed for clarity. Here, the extended corona emission has $PA$ perpendicular to the disk surface locating in the 4th quadrant (purple lines), while the $PA$ for the power-law component from ${\it INTEGRAL}$ SPI observations lies in the 1st quadrant (dashed blue lines).
The possible vector lengths (i.e., $PF$ values) reach their maximum values when the angle between the two vectors is maximized (maximum cancellation). Following Fig.~\ref{fig:QU_extend}, this happens for the corona contribution line~A and jet contribution line~B. 
$PF_{\rm corona}$ is calculated as $\sim$2.5\% (vector~C \textcolor{black}{with \mbox{($Q/I$, $U/I$) = (0.015, -0.020)} has length 0.025 following Eq.~\ref{eq:PF}}) from the numerical simulation \citep{sim_extend}, in which case $PF_{\rm jet}$ will be limited to \mbox{$<$5\%} (vector~D \textcolor{black}{with \mbox{($Q/I$, $U/I$) = (0, 0.050)} has length 0.05)}, which is where the sum of the two vectors intersects the ${\it PoGO+}$ upper limit.

If we assume that \mbox{$PF_{\rm jet} <5\%$} from the ${\it PoGO+}$ results and that the the jet emission is polarized \mbox{$(76\pm15)$\%} from \citet{integral_spi}, then the jet flux will be less than about 8\% of the total flux (i.e., $F_{\rm jet} < 2 \times 10^{-9}$ erg s$^{-1}$ cm$^{-2}$ in the 20--180~keV range, following from
$F_{\rm jet}$ $<$ 0.05 $\times$ $F_{\rm {total}} \times (1/0.76)$).

From the SED of \mbox{Cyg X-1}, \citet{cygx1_sed} estimate the emission of the synchrotron jet component for two cases: several 100~keV flux dominated by the jet, or the non-thermal corona emission.
The jet-dominated case corresponds to the above picture for the additional highly-polarized power-law component. The 20--180~keV flux $F_{\rm jet}$ then becomes \mbox{9 $\times$ 10$^{-10}$ erg s$^{-1}$ cm$^{-2}$}, where we apply power-law emission with a photon index of~1.6 and a normalization of \mbox{0.05 photons s$^{-1}$ cm$^{-2}$ keV$^{-1}$} at 1 keV.
This prediction is below the current upper limit resulting from the ${\it PoGO+}$ measurements, meaning we cannot distinguish between \textcolor{black}{jet and non-thermal corona emission} dominating the several 100~keV flux.

\subsection{Lamp-post corona model with an additional component}
\label{sec:lamp}

\begin{figure}
    \includegraphics[width=\columnwidth]{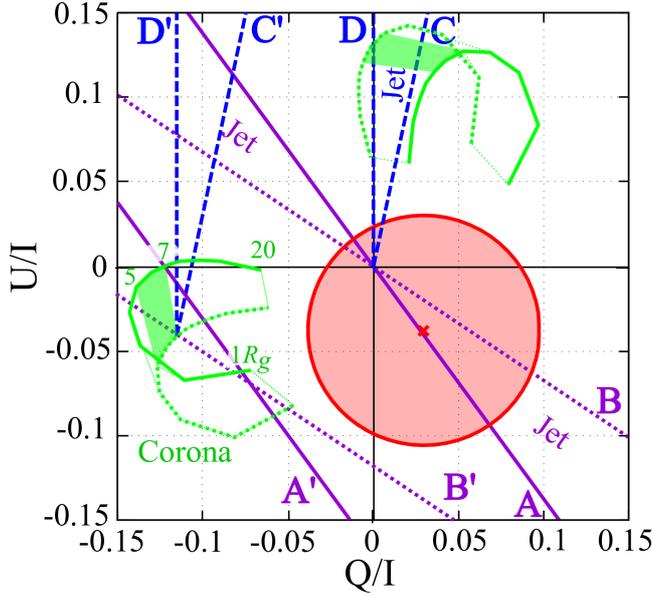}
	\caption{
    Vector calculations for the 
    \textcolor{black}{simple}
    lamp-post corona case
    \textcolor{black}{with a compact corona height of 5-7 $R_g$}.
    The lamp-post corona model results \textcolor{black}{in} \mbox{$PF$ $\sim$ 13\%} and $PA$ shifted by $\sim$60$^{\circ}$ (i.e. $\sim$120$^{\circ}$ in the $QU$ plane) relative to the disk rotation axis, 
    resulting from spectral analyses and simulations \citep{cygx1_fabian, sim_lamp} assuming CW disk rotation \textcolor{black}{(filled green region in the 3rd quadrant) and CCW disk rotation (filled green region in the 1st quadrant)}.
    \textcolor{black}{Lines A' and B'}
    {are parallel with lines A and B which describe $PA_{\rm jet}$ perpendicular/parallel to the disk surface.
    Solid and dotted lines correspond to the jet direction uncertainty of $-5^{\circ}$ and $+5^{\circ}$, respectively.
    Lines C' and D' are parallel with lines C and D which describe $PA_{\rm jet}$ arising from the power-law component by ${\it INTEGRAL}$.
    }
    }
    \label{fig:QU_lamp}
\end{figure}

We now re-visit the lamp-post corona model, since in the presence of an additional emission component, $PF_{\rm corona}$ can become higher than $PF_{\rm total}$, 
\textcolor{black}{through cancellation by $PF_{\rm jet}$ perpendicular/parallel to the disk surface or $PF_{\rm jet}$ from the power-law component by ${\it INTEGRAL}$}.
\textcolor{black}{We first consider the simple lamp-post corona model
with the compact corona height of 5--7~$R_{\rm g}$ estimated from spectral analyses \citep{cygx1_fabian},}
\textcolor{black}{corresponding to the two filled green regions in Fig.~\ref{fig:QU_lamp} for CW and CCW disk rotation.}
\textcolor{black}{Auxiliary lines (A', B', C', D') have been added parallel with their counterparts (A, B, C, D), with magenta lines (A, A', B, B') corresponding to $PA_{\rm jet}$ perpendicular/parallel to the disk surface and blue lines (C, C', D, D') to $PA_{\rm jet}$ from the power-law component by ${\it INTEGRAL}$.}
\textcolor{black}{For clarity, only lines for CW disk rotation, as from radio observations \citep{cygx1_inc_radio}, have been drawn (3rd quadrant) although conclusions do not change if the disk rotation is instead CCW (1st quadrant).}
\textcolor{black}{
Lines A', B' C', D' intersect the filled green region at the side corresponding to corona height 7 $R_g$, where they come closest to the PoGO+ region.
Lines A' and B' are based on $\pm 5^{\circ}$ uncertainty in the radio jet direction, where solid and dotted lines correspond to the uncertainty in the negative and positive direction, respectively.
Then, dotted line B' ($+5^{\circ}$ uncertainty) cannot intersect at 7 $R_g$ of the top solid green line ($-5^{\circ}$ uncertainty).
Lines C' and D' from {\it INTEGRAL} observations are independent from the radio jet, and can intersect the filled green region anywhere.
}
None of these lines cross the 90\% upper-limit region of the ${\it PoGO+}$ measurement, 
\textcolor{black}{i.e., the jet $PA$ required to match the $PoGO+$ data is inconsistent with the three jet directions considered here: perpendicular to the disk, parallel to the disk, and direction as suggested by ${\it INTEGRAL}$ results.}
Therefore, we conclude that the \textcolor{black}{simple} lamp-post corona model cannot explain the observational results for \mbox{Cyg X-1}, even when considering an additional emission component.


\textcolor{black}{While a simple lamp-post corona is excluded, there may be more complex cases which this study cannot rule out, e.g. if the corona is outflowing or elongated.
If a corona height is close to 1 or \mbox{20 $R_g$}, which is inconsistent with previous spectral analyses (5--7 $R_g$), it could match the ${\it PoGO+}$ upper limit in the presence of a jet with PA parallel to {\it INTEGRAL} power-law component or perpendicular to the disk surface, respectively.
More detailed simulations would be required to study such cases.}

\section{Conclusions}
\label{sec:conclusion}

We have studied the ${\it PoGO+}$ hard X-ray polarization results for the BHB \mbox{Cygnus X-1} in the Cartesian Stokes $QU$ plane.
When only emission from the corona is considered,
the extended corona model (low $PF$ and $PA$ perpendicular to the disk surface) is consistent with the ${\it PoGO+}$ 90\% upper limit,
while the \textcolor{black}{simple} lamp-post corona model (high $PF$ and rotated $PA$ values) does not match our observations, reaffirming results from our previous paper \citep{PoGO+_CygX1}.

For corona emission together with a possible synchrotron jet  component, we use results in the Stokes $QU$ plane to estimate the upper limit of the jet flux.
When assuming a typical $PF$ value of a few percent for the extended corona, the remaining $PF_{\rm jet}$ can be \mbox{$<$5--10\%} for $PA$ either perpendicular to the disk surface, similar to infrared and optical \citep{cygx1_ir_opt}, or \mbox{$PA$ $\sim$40$^{\circ}$}, as proposed from ${\it INTEGRAL}$ data in the several 100~keV range \citep{integral_spi}.
The upper flux limit of a highly-polarized jet component is estimated as \mbox{$F_{\rm jet}$ $<$ (2--3) $\times$ 10$^{-9}$ erg s$^{-1}$ cm$^{-2}$}.

Although $PF_{\rm corona}$ in the lamp-post corona model can be higher through cancellation by a possible component $PF_{\rm jet}$, the predicted $PF_{\rm corona}$ value is still too high to explain the ${\it PoGO+}$ results for \mbox{Cyg X-1}.

Current and near-future X-ray polarization missions such as {\it X-Calibur} \citep{xcalibur} (balloon-borne), {\it AstroSat} \citep{astrosat} and {\it IXPE} \citep{ixpe} (satellites) can further constrain the jet emission of \mbox{Cyg X-1}. 
For this, simulations specific to \mbox{Cyg X-1} (inclination angle, truncated disk radius, etc.) will be required, since current simulations are for generic cases \citep{sim_extend}.
Next-generation gamma-ray missions such as {\it e-ASTROGAM} \citep{eastrogam} and {\it AMEGO} are designed with polarimetric capabilities. These will directly confirm the polarization information above several 100~keV, allowing the jet contribution to be determined.


\section*{Acknowledgements}

This research was supported by The Swedish National Space Agency, The Knut and Alice Wallenberg Foundation, The Swedish Research Council, The Japan Society for Promotion of Science, and ISAS/JAXA.








%
%


\bsp	
\label{lastpage}
\end{document}